# CMOS-compatible controlled hyperdoping of silicon nanowires

Yonder Berencén,[a,*] Slawomir Prucnal,[a] Wolfhard Möller,[a] René Hübner,[a] Lars Rebohle,[a] Roman Böttger,[a] Markus Glaser,[b] Tommy Schönherr,[a] Ye Yuan,[a] Mao Wang,[a] Yordan M. Georgiev,[a,§] Artur Erbe,[a] Alois Lugstein,[b] Manfred Helm,[a,c] Shengqiang Zhou,[a] and Wolfgang Skorupa[a]

[a]Helmholtz-Zentrum Dresden-Rossendorf, Institute of Ion Beam Physics and Materials Research, Bautzner Landstraße 400, D-01328 Dresden, Germany

[b]Institute for Solid State Electronics, Vienna University of Technology, Floragasse 7, A-1040 Vienna, Austria

[c]Center for Advancing Electronics Dresden (cfaed), Technische Universität Dresden, 01062 Dresden, Germany

*Corresponding author: y.berencen@hzdr.de

§On leave of absence from the Institute of Electronics at the Bulgarian Academy of Sciences, Sofia, Bulgaria

**Abstract:**

Hyperdoping consists of the intentional introduction of deep-level dopants into a semiconductor in excess of equilibrium concentrations. This causes a broadening of dopant energy levels into an intermediate band between the valence and conduction bands.[1,2] Recently, bulk Si hyperdoped with chalcogens or transition metals has been demonstrated to be an appropriate intermediate-band material for Si-based short-wavelength infrared photodetectors.[3-5] Intermediate-band nanowires could potentially be used instead of bulk materials to overcome the Shockley-Queisser limit and to improve efficiency in solar cells,[6-9] but fundamental scientific questions in hyperdoping Si nanowires require experimental verification. The development of a method for obtaining controlled hyperdoping levels at the nanoscale concomitant with the electrical activation of dopants is, therefore, vital to understanding these issues. Here, we show a CMOS-compatible technique based on non-equilibrium processing for the controlled doping of Si at the nanoscale with dopant concentrations several orders of magnitude greater than the equilibrium solid solubility. Through the nanoscale spatially controlled implantation of dopants, and a bottom-up template-assisted solid phase recrystallization of the nanowires with the use of millisecond-flash lamp annealing, we form Se-hyperdoped Si/SiO$_2$ core/shell nanowires that have a room-temperature sub-band gap optoelectronic photoresponse when configured as a photoconductor device.

## 1. Introduction

The need of using nanostructured materials or nanostructures such as quantum dots or nanowires (NWs) for applications including photovoltaics, nanoelectronics, neuroscience, plasmonics, sensing and optoelectronics is nowadays undisputed.[7-16] The large surface area to volume ratio of nanomaterials provides further benefits as compared to bulk or thin films. Yet the success of these materials ultimately depends on the possibilities of controlling their properties through doping.

While different approaches for doping Si NWs have been explored,[17-21] hyperdoping remains elusive but might enable NWs with unique electrical and optical properties such as metal-insulator transition,[2] sub-bandgap optical absorption[3] and localized surface plasmon resonances.[12] However, the possibility of hyperdoping Si NWs in the same manner as bulk Si, the type of microstructural defects and electronic states formed in the NWs due to hyperdoping, the distribution of dopants and charge carriers as well as the resulting optical and transport properties have yet to be addressed experimentally.

Unlike traditional dopants such as B and P, Se is the dopant of choice, because it introduces deep donor levels in Si (viz. from around 0.3 to 0.6 eV) that broaden into an intermediate band when Se is present at concentrations greater than the equilibrium solid solubility.[2,3] The higher the Se concentration, the broader the intermediate band, which ultimately merges with the conduction band. This leads to a strong delocalization of electrons into the intermediate band that provides carriers. By using a doping technique that is standard in the semiconductor manufacturing industry, viz. ion implantation and flash lamp annealing (FLA), hyperdoped Si with Se has recently been achieved in the bulk.[22] Alternatively, B- and P-hyperdoped Si nanocrystals have already been realized via kinetics-controlled mechanisms such as nonthermal plasma.[23] In contrast to ion implantation, the nonthermal plasma, which relies on the probability of collision between a Si nanocrystal and a dopant atom and the binding energy of dopants at the nanocrystal surface, suffers from precise control over the total areal dose of the dopants.

Hyperdoped Si is a thermodynamically metastable material that can only be obtained using non-equilibrium methods. Therefore, FLA is a suitable annealing method due to its short time annealing scale of milliseconds or less.[24] We performed temperature simulations using COMSOL Multiphysics© software which revealed that the surface temperature achieves a maximum of approximately 1230 K during the flash. Detailed investigations devoted to light absorption in Si NWs on $SiO_2$ show that there is a strong absorption in the visible and the UV spectral region for Si NWs with a diameter on the order of 100 nm and greater[28] and we would expect similar absorption for Si/$SiO_2$ core/shell NWs. Considering that the flash lamp spectrum has a maximum in the blue spectral region,[24] most of the light, which is incident on the NW, is then absorbed.

The crystallization of the NWs is more complex compared to the bulk case. Although the crystallization is an atomistic process,[26] interfaces affect the process and the dominance of surfaces in nanostructures gives a significant overall contribution. It is known from the regrowth of amorphous fins in field effect transistors during rapid thermal annealing that the growth at the Si-$SiO_2$ interfaces is delayed, which finally favors the formation of a partly polycrystalline NW.[27]

Possible reasons for this delay are the greater stability of bond defects at the surface[28] and the dependence of the crystallization velocity on the crystal orientation.[29]

Here, we report on non-equilibrium processing for controlled hyperdoping of Si/SiO$_2$ core/shell nanowires previously synthesized by the vapor-liquid-solid method and plasma-enhanced chemical vapor deposition. Our approach is based on Se implantation into the upper half of Si/SiO$_2$ core/shell NWs followed by millisecond-flash lamp annealing, which allows for a bottom-up template-assisted recrystallization via solid phase of the amorphized parts of the Si/SiO$_2$ core/shell NWs upon high-fluence Se implantation. The Se-hyperdoped Si/SiO$_2$ core/shell NWs are recrystallized and accommodate Se concentrations as high as $10^{21}$ cm$^{-3}$. As a proof of device concept, we demonstrate a single Se-hyperdoped NW-based IR photoconductor. These results establish the combination of ion implantation and flash lamp annealing as a promising nanoscale hyperdoping technology.

## 2. Results and Discussion

We used Si NWs grown using the vapor-liquid-solid method.[30] After growing, a 20 nm thick SiO$_2$ shell was formed on the NWs through plasma-enhanced chemical vapor deposition, to protect the integrity of the NWs during ion implantation. The Si/SiO$_2$ core/shell NWs were transferred onto a 70-nm thick SiO$_2$ layer deposited on a Si substrate. The transferred NWs were then implanted with Se ions at a fluence of $1\times10^{16}$ cm$^{-2}$ or $3\times10^{16}$ cm$^{-2}$ (**Figure 1a**) and an implantation energy of 60 keV to ensure that the peak Se concentration was in the upper half of the NW core. Whereas the upper half of the Si core is amorphized during high-fluence implantation, the bottom half is left undamaged to be used as a seed for recrystallization during the annealing stage. The choice of the Se implantation fluences was based on the range of atomic Se concentrations (1 to 2%), where the intermediate band has been reported to be formed in the bulk Si.[22] These implantation conditions led to non-equilibrium Se concentrations as high as $10^{21}$ cm$^{-3}$ in the implanted region, which is five orders of magnitude greater than the equilibrium solid solubility of Se in Si ($10^{16}$ cm$^{-3}$).[31] To be electrically activated, the Se dopant atoms must migrate from interstitial positions to substitutional sites in the Si lattice. We achieved this through FLA (see Experimental Section) of the implanted Si/SiO$_2$ core/shell NWs (**Figure 1b**).

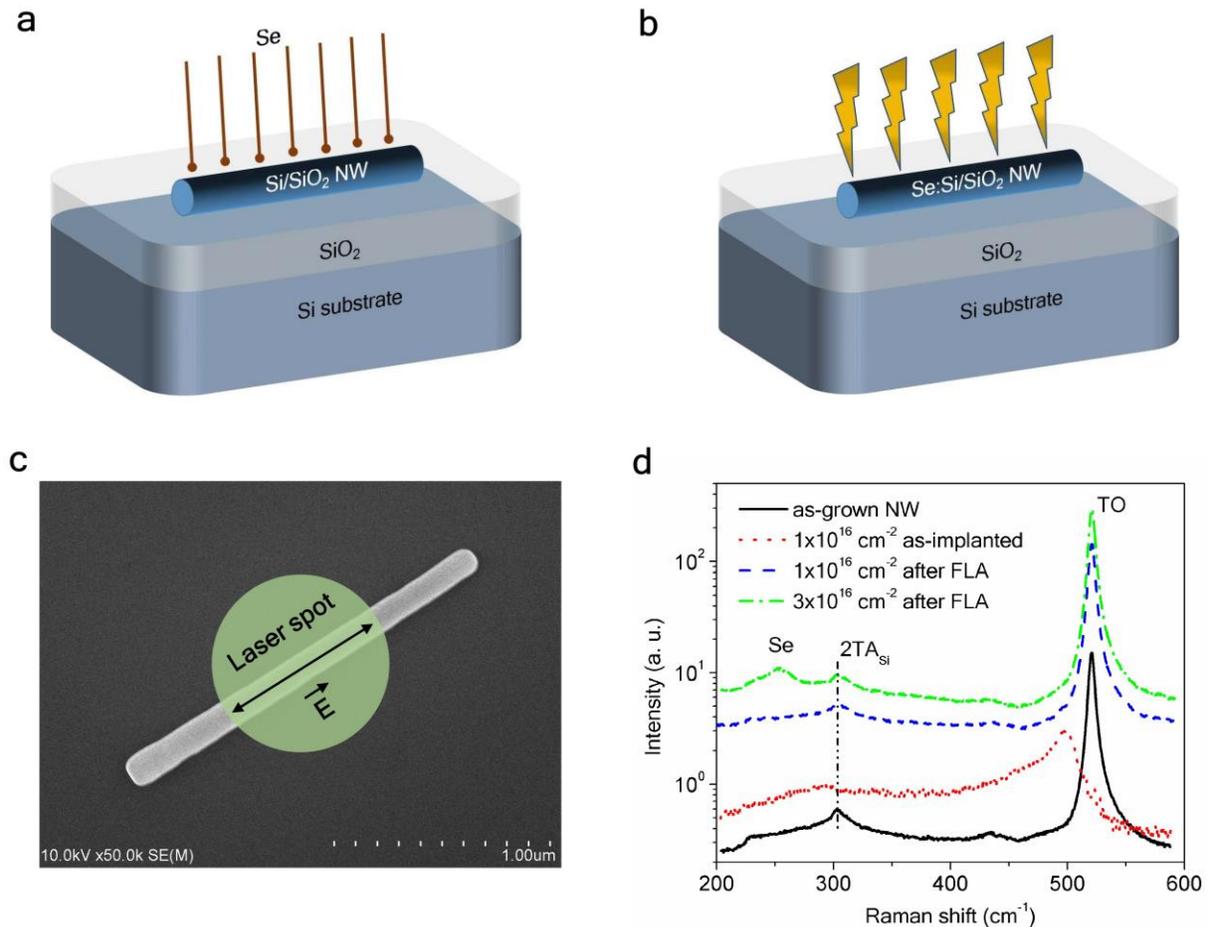

**Figure 1.** Schematic illustration of the ion beam hyperdoping process of individual Si/SiO$_2$ core/shell nanowires and their crystalline properties after different stages of hyperdoping. (a) Room-temperature Se ion implantation of individual Si/SiO$_2$ core/shell NWs in which the peak dopant concentration was targeted to the upper half of NWs. (b) Millisecond-flash lamp annealing ensures that the annealing temperatures are less than the melting point of amorphous Si (approx.

1480 K)[35]. This allows for a bottom-up template-assisted solid phase recrystallization of the hyperdoped NW. (c) Top view of an SEM image of a single NW after ion implantation followed by FLA, with the laser polarization for the Raman characterization of individual NWs indicated as $\vec{E}$. (d) Room-temperature μ-Raman spectra of as-grown crystalline Si/SiO$_2$ core/shell NW as a reference (black solid line), $1\times10^{16}$ Se/cm$^2$-implanted Si/SiO$_2$ core/shell NW before (red dotted line) and after FLA (blue dashed line), and $3\times10^{16}$ Se/cm$^2$-implanted Si/SiO$_2$ core/shell NW after FLA (green dashed-dot line).

We used μ-Raman spectroscopy to investigate the crystalline properties of individual Si/SiO$_2$ core/shell NWs after different stages of hyperdoping. To this end, NWs were transferred onto Scotch tape, where the linearly polarized laser light was aligned with the long axis of the NW (see Experimental Section) as illustrated in a typical scanning electron microscopy image (**Figure 1c**).

Consequently, we show the Raman scattering spectra of the individual as-grown single crystalline, the as-implanted and the flash-lamp annealed Si/SiO$_2$ core/shell NWs (**Figure 1d**). The phonon modes of bulk crystalline Si are well known to be located at 520 cm$^{-1}$ and 303 cm$^{-1}$, representing the first-order optical phonon (TO) and the second-order two transverse acoustic phonon (2TASi) scattering, respectively.[32] In the as-implanted Si NW, the broad band peak at approximately 480 cm$^{-1}$ is evidence of a strong amorphization[33] of the upper part of the Si NW core that correlates with the high-fluence Se implantation. By contrast, a peak superimposed to the broad band and close to 497 cm$^{-1}$ is also observed, which corresponds to Si-Si vibrational modes at the end of the amorphization zone[33] viz. to end-of-range defects formed upon ion implantation. After FLA, however, the crystalline structure of the hyperdoped Si/SiO$_2$ core/shell NWs at both Se fluences is fully restored without traces of the amorphous band, as seen in **Figure 1d**. No peak shift is observed in either the TO or the 2TASi modes for the flash-lamp annealed NWs if compared with the as-grown single crystalline Si NWs indicating that the hyperdoped NWs were structurally relaxed. In NWs produced under the greater Se fluence of $3\times10^{16}$ cm$^{-2}$, we also identified a phonon mode at approximately 250 cm$^{-1}$ (**Figure 1d**), which is related to a longitudinal mode of substitutional Se dimers in the crystalline Si.[34] Hereafter, we only focus on $1\times10^{16}$ Se/cm$^2$-implanted Si/SiO$_2$ core/shell NWs because Se-Se dimers are electrically inactive complexes that strongly affect to both the electrical and optical activation of the Se dopants.[3]

To inspect the element distribution within the Si/SiO$_2$ core/shell NWs, we used high-angle annular dark-field scanning transmission electron microscopy (HAADF-STEM) in conjunction with energy-dispersive X-ray spectroscopy (EDXS) element mapping. Interestingly, relative to as-implanted NWs (**Figure 2a**) and Se-hyperdoped bulk Si annealed by nanosecond-pulsed laser irradiation,[22] we found no observable redistribution of Se atoms following FLA (**Figure 2b-c**). We also detected that either longer FLA times or preheating temperatures above 875 K are sufficient to induce an unwanted out-diffusion of Se atoms towards the Si/SiO$_2$ core/shell interface and thus their electrical deactivation. In addition, we propose that the SiO$_2$ shell prevented ion sputtering and/or deformation of the Si/SiO$_2$ core/shell NWs during high-fluence ion implantation, in contrast to the as-grown unprotected Si NWs, which are subjected to ion irradiation.[36] Therefore, these results clearly indicate that (1) the dopant profile is mostly determined by the ion implantation step and (2) FLA in the millisecond range is able to fully restore the crystalline structure of the hyperdoped Si/SiO$_2$ core/shell NWs (as shown in **Figure 1d**) without causing dopant diffusion.

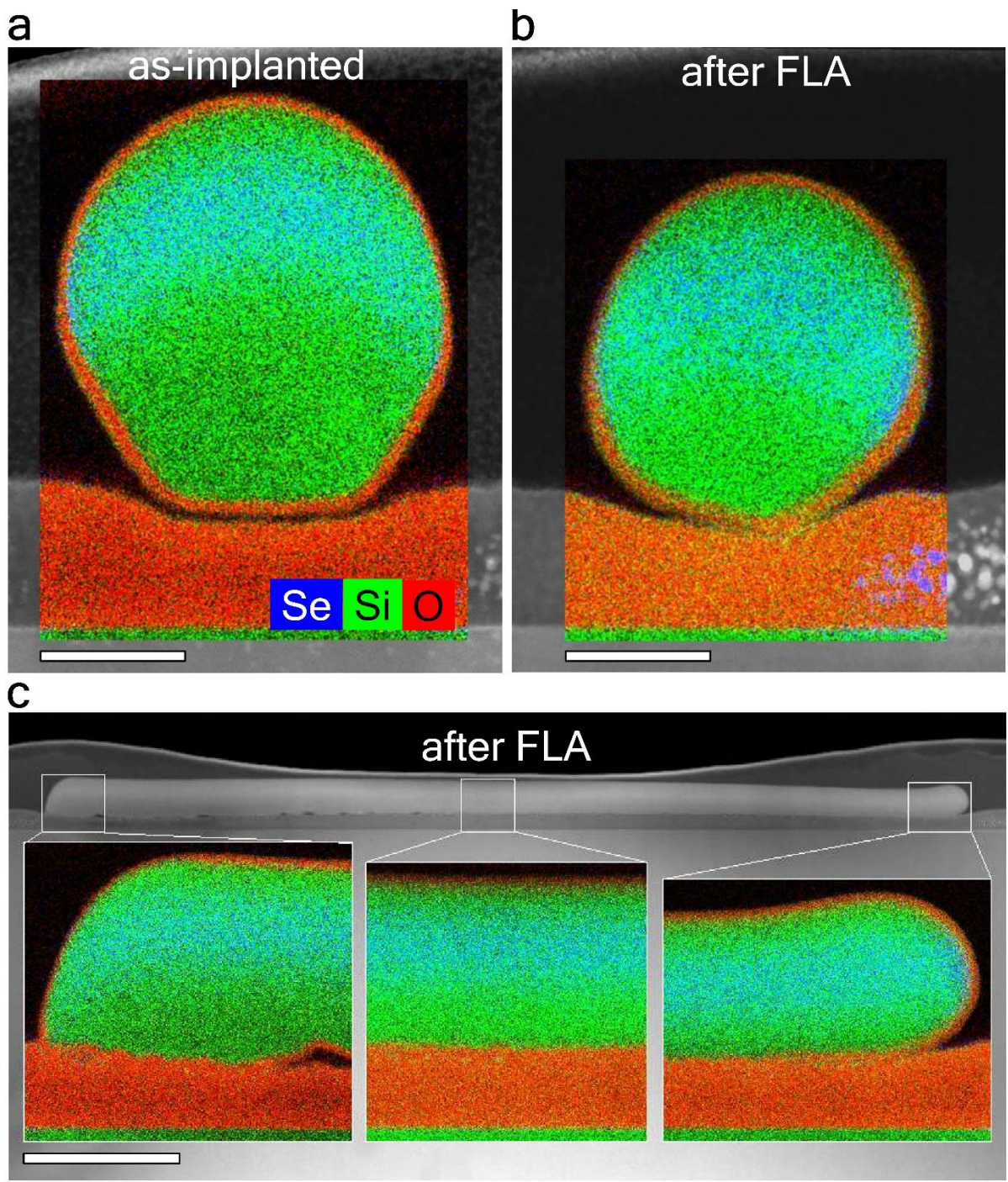

**Figure 2.** Spatially-resolved chemical analysis. Cross-sectional HAADF-STEM images superimposed with EDXS element maps for (a) a $1\times10^{16}$ Se/cm$^2$-implanted Si/SiO$_2$ core/shell NW

and (b), a comparable NW after subsequent FLA. (c) Longitudinal element distribution of a comparable NW after FLA; scale bars: 50 nm in a, b and 500 nm in c.

To investigate the high-fluence ion implantation process at nanoscale that gives rise to Se-hyperdoped Si/SiO$_2$ core/shell NWs, we performed computational simulations using a three-dimensional dynamic Monte Carlo TRI3DYN code[37] (see Supplementary Information). **Figure 3** presents representative cross-sectional views of the different computational simulations performed on the Si/SiO$_2$ core/shell (110 nm core/20 nm shell) nanowire resting on a SiO$_2$ layer deposited atop a Si substrate. The implanted Se atoms are evenly distributed on the upper half of the NW (**Figure 3a**). The shape of the NW core is also preserved by the SiO$_2$ shell, which is partially sputtered during the high-fluence Se implantation. Neither atom migration nor changes in the morphology of the NWs were observed during implantation. **Figure 3b** depicts a quantitative analysis of the Se concentration derived from the dynamic Monte Carlo computer simulations for implantation energies of 60 keV. Se concentrations up to the order of $10^{21}$ Se/cm$^3$ were found in the upper half of the NW core. This is in line with our experimental results (**Figure 2**), and confirms that the simulations reproduced the desired doping concentrations and distributions. In addition, we computationally quantified, through the use of the critical point-defect-density model[38] the ion-induced damage in the NWs in terms of the number of displacements per target atom, dpa. In **Figure 3c**, we use black to show the transition zone between the amorphized region and the crystalline one, which remains unaffected after the high-fluence Se irradiation. This indicates that the adequate choice of implantation parameters together with the thickness of the SiO$_2$ shell allows for retaining the bottom part of the NW free of the ion-induced damage.

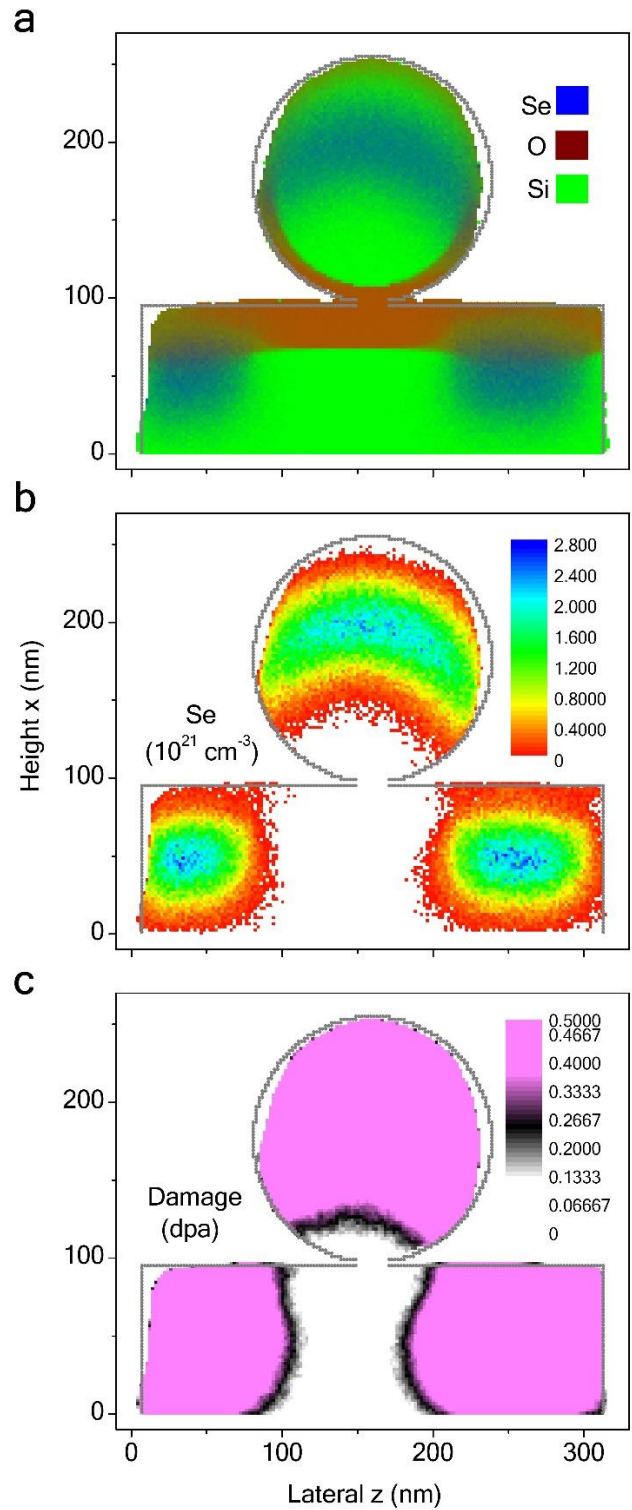

**Figure 3.** Dynamic Monte Carlo computer simulations of high-fluence Se-implanted Si/SiO$_2$ core/shell NWs at 60 keV. (a) Atomic fractions of O (brown), Si (green) and Se (blue) at a fluence

of 1×10$^{16}$ Se/cm$^2$; Se is scaled ×20. (b) Se atomic density at 1×10$^{16}$ Se/cm$^2$. (c) Structural damage computed after high-fluence Se implantation given in terms of displacements per atom (dpa), where the dark contour indicates the transition zone between the amorphized region and the still crystalline region based on the critical point-defect-density model. The overall gray contours stand for the original shape of the NW.

Next, to examine the specific resistivity of individual hyperdoped Si/SiO$_2$ core/shell NWs, we therefore fabricated a four-probe NW structure[36] (**Figure 4a**). This value was measured to be 1.27 Ω·cm and is comparable to that of boron-doped Si NWs with dopant concentrations in excess of 10$^{19}$ cm$^{-3}$.[39] Subsequently, to characterize the electrical activation of Se dopants after the recrystallization step using FLA, low-temperature three-probe NW Hall effect experiments with nanoscale spatial resolution[40] were performed. The electrically active Se concentration in NWs implanted with 1×10$^{16}$ Se/cm$^2$ was determined to be 3.3×10$^{20}$ cm$^{-3}$ at 2.3 K (see Experimental Section). This results in an electrical activation efficiency of Se dopants of approximately 30%, which is comparable with Se-hyperdoped bulk Si fabricated through the ion implantation followed by FLA.[3] Indeed, the measured carrier concentration in Se-hyperdoped Si/SiO$_2$ core/shell NWs is three orders of magnitude greater than that of P-doped Si NWs developed by ion implantation at an identical fluence of phosphorous (1×10$^{16}$ P/cm$^2$).[21] The electrically inactive Se atoms are likely due to the formation of clusters and/or interstitials.[41]

Having established that, we then fabricated photoconductor devices based on individual NWs to verify the sub-bandgap photoresponse that is thought to be mediated by the formation of the intermediate band in the upper half of the Si NW bandgap. **Figure 4b** schematically illustrates the photoconductor device, in which photons with an energy of less than the 1.12 eV Si bandgap are expected to be absorbed within the NW because of the large concentration of the mid-gap dopant levels created by hyperdoping with the Se impurities. The absorbed photons are then converted into a measurable electrical current at one of the electrical contacts. The inset in **Figure 4c** depicts a top view image captured with an optical microscope of a typical single Si NW-based IR photoconductor device. For the sake of brevity and to be consistent with the dimensions of the investigated NW in **Figure 2c**, we show only the results of single Si NW-based IR photoconductor devices with a fixed nanowire diameter of approximately 130 nm (including a 20 nm shell) and a spacing between electrodes of approximately 2 µm. We then conducted room-temperature current-voltage (I-V) measurements in dark conditions and in response to a 1.55 µm (0.8 eV) laser light at 35 mW. For a bare Si/$SiO_2$ core/shell NW, we found no difference between I-V curves obtained in the dark and under illumination (**Figure 4c**). By contrast, for Se-hyperdoped Si/$SiO_2$ core/shell NWs, we observed (1) an enhancement of the electrical conductivity in dark conditions compared with the bare Si/$SiO_2$ core/shell NW and (2) a detectable increase in the conductivity at a 1.55 µm illumination. These two aspects experimentally demonstrate the presence of the intermediate band in the Si/$SiO_2$ core/shell NW bandgap related to Se hyperdoping and the electrical and optical activation of the Se impurities.

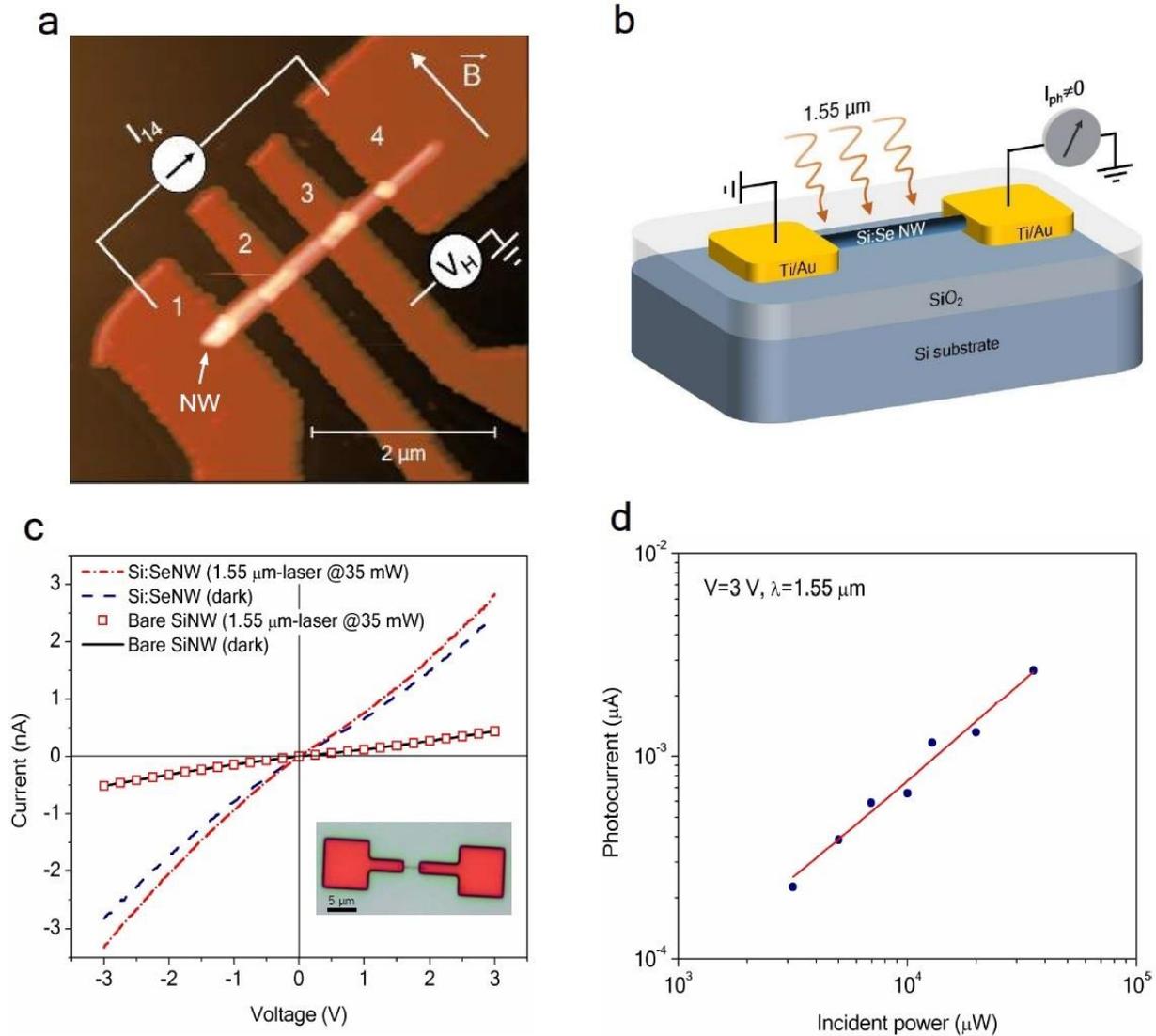

**Figure 4.** Single Si NW-based IR photoconductor. (a) Atomic force microscopy image of a four-probe NW structure for resistivity measurements. Probes or contacts are denoted as 1, 2, 3 and 4. The two inner contacts allow for spatial resolution in the Hall measurements using the three-probe NW Hall configuration. (b) Three-dimensional schematic view of the single Se-hyperdoped Si/SiO$_2$ core/shell NW photoconductor and the 1.55-μm-focused laser used to illuminate the device. (c) Typical room-temperature current-voltage characteristic under dark and 1.55-μm illumination of a single bare Si/SiO$_2$ core/shell NW and a $1\times10^{16}$ Se/cm$^2$-hyperdoped Si/SiO$_2$

core/shell NW, both with diameters of approximately 130 nm. The inset shows an optical microscope image of an NW, and its contacts created by electron beam lithography. (d) Photocurrent as a function of incident optical power at a wavelength of 1.55 μm. Please note that the laser spot is much larger than the footprint of the NW, which makes the estimation of the device efficiency difficult.

Next, we recorded the intensity of the photocurrent (calculated as $I_{light}$ - $I_{dark}$) measured at an applied bias of 3 V as a function of the incident optical power at 1.55 μm. **Figure 4d** reveals that the photocurrent increases linearly with the incident optical power, which suggests a single-photon absorption process mediated by the intermediate band. This type of mechanism is in good agreement with room-temperature photodetectors that make use of Se- and Au-hyperdoped bulk Si as the intermediate band material.[3,4] We also found no significant differences on the hyperdoping process and the properties of the hyperdoped Si/SiO$_2$ core/shell NWs within the range of investigated NW diameters (from 80 nm to 160 nm).

## 3. Conclusion

In conclusion, we have established a method for the hyperdoping of semiconductors at the nanoscale, as applied to Si/SiO$_2$ core/shell NWs. The proposed method is fully compatible with the semiconductor industry, and can be further extended to other types of semiconductors with nanometer-scale dimensions and both traditional and unconventional dopants in a wide range of dopant concentrations with an appropriate choice of fabrication parameters. This latter especially applies to FLA that has to consider the different thermodynamic properties of materials, such as melting temperature, crystallization behavior, diffusion, etc.

## 4. Experimental Section

*Fabrication of Si/SiO₂ core/shell NWs*: Nanowires were epitaxially grown on <111> Si substrates via the vapor-liquid-solid growth mechanism in a conventional low-pressure chemical vapor deposition system. Silane (SiH$_4$) was used as a precursor gas and a thin magnetron-sputtered Au layer (<3 nm) was used as a catalyst. After NW growth, the Au was removed with aqua regia. The resulting as-grown NWs are <111> oriented with diameters ranging from 80 to 160 nm and an average length of approximately 7 μm. Further details about the growing of NWs can be found elsewhere.[30] The NWs were then subjected to plasma-enhanced chemical vapor deposition to form a 20-nm thick SiO$_2$ shell. Subsequently, the Si/SiO$_2$ core/shell NWs were transferred through sonication in ethanol onto a 70 nm thick SiO$_2$ layer deposited by low-pressure chemical vapor deposition on top of a Si substrate.

*Ion beam implantation*: Se implantation was performed using the 500-kV ion implanter at the Ion Beam Center at Helmholtz-Zentrum Dresden-Rossendorf. Ion beams were extracted from an IHC Bernas ion source with solid Se source feed, subsequently mass-separated by a sector magnet, focused by an electrostatic Einzel lens, and raster scanned over the sample area using two electrostatic plates driven by a high voltage triangle signal at a frequency of 1 kHz. Additionally, the ion beam was deflected by 15° just before the raster scanning by another pair of electrostatic plates to avoid neutral particle contamination in the ion beam. The implanted ion fluence was measured using four corner Faraday cups, with secondary electron suppression, in front of the sample plane, which were connected to a calibrated charge integrator. The pressure in the beam line was $5 \times 10^{-7}$ mbar, while the pressure in the implantation chamber was kept below $2.5 \times 10^{-6}$ mbar.

*Non-equilibrium thermal treatment*: Se-implanted Si/SiO$_2$ core/shell NWs were flash-lamp annealed at an energy of 33 J/cm$^2$ for 1.3 ms in N$_2$ atmosphere. A preheating step was also performed at 663 K after 30 s of a temperature ramp to reduce the temperature gradient in the silicon substrate during the FLA stage. Further details about the FLA system can be found elsewhere.[24]

*Structural characterization of individual hyperdoped Si/SiO$_2$ core/shell NWs:* The phonon spectra were determined using µ-Raman spectroscopy of individual NWs that were transferred onto Scotch tape, which we used as a dummy substrate. This eliminated phonon contributions from the bare Si substrate. Spectra were obtained in the wavenumber range of 200 to 600 cm$^{-1}$ with a resolution of approximately 0.5 cm$^{-1}$. The Raman scattering was excited with a 532 nm Nd:YAG laser, linearly polarized along the axis of the NW, to optimize Raman signal intensity. The power of the laser was kept sufficiently low to avoid any shift in the phonon peaks arising from laser-induced heating. A 100× objective lens was used to collect the backscattered light that was then detected using a cryogenically cooled charge-coupled device camera attached to a spectrometer. All the measured phonon spectra were corrected by the Raman spectrum of the Scotch tape.

*Spatially-resolved chemical analysis:* HAADF-STEM imaging and spectrum imaging based on EDXS were performed at 200 kV with a Talos F200X microscope equipped with an X-FEG electron source and a Super-X EDXS detector system (FEI). Prior to STEM analysis, the specimen was mounted in a high-visibility low-background holder and was placed for 10 s into a Model 1020 plasma cleaner (Fischione) to remove contamination. TEM lamellae of NW cross and longitudinal sections were prepared by in situ lift-out using a Zeiss Crossbeam NVision 40 system. To protect the NW surface at the area of interest, a carbon cap layer was first deposited aided by an electron beam and followed by Ga-focused ion beam (FIB) assisted precursor decomposition.

Subsequently, the TEM lamella was prepared using a 30 keV Ga FIB with adapted currents and then transferred to a 3-post copper lift-out grid (Omniprobe) using a Kleindiek micromanipulator. To minimize sidewall damage, Ga ions with 5 keV of energy were used for final thinning of the TEM lamella to electron transparency.

*Device fabrication*: The devices were fabricated by creating two electrical contacts on opposite sides of the Si/SiO$_2$ core/shell NWs supported by a previously oxidized Si substrate. In the first step, an array of markers was written onto the sample using electron beam lithography (EBL) with an e_LiNE plus system (Raith GmbH). These marker structures were then used to accurately measure the position of single NWs using the SEM mode of the same EBL system. In the third step, contact pads connected to the ends of individual wires were written again by EBL. To ensure good contact quality, reactive ion etching was carried out to remove the SiO$_2$ shell at the contact parts just before the Ti/Au metallization step performed using an electron beam metal evaporation tool. Finally, a standard lift-off process using acetone and isopropanol was applied. The same routine was used to fabricate the four-probe NW structures for the resistivity and spatially-resolved Hall effect measurements.

*Resistivity and Hall effect measurements on individual hyperdoped Si/SiO$_2$ core/shell NWs*: Room-temperature resistivity measurements were performed on a four-probe NW structure. The value was estimated using the formula $\rho_s = V_{23}\pi r^2/LI_{14}$, where $V_{23}$ is the voltage drop across the contacts 2 and 3, $I_{14}$ is the constant current through contacts 1 and 4, $r$ is the NW core radius and $L$ is the gap between the contacts 2 and 3. Next, Hall effect measurements were executed following the three-probe NW Hall configuration proposed by Hultin *et al.*[40] In detail, a current of 10 nA was applied through the NW under an in-plane magnetic field swept from -2 T to 2 T at 2.3 K with the help of a Lakeshore system. The Hall voltage was measured in one of the inner contacts as a voltage change referred to ground (see **Figure 4a**). This change in the voltage is driven by the variations in the applied magnetic field that is perpendicularly aligned with the NW. To separate the Hall voltage from drift effects caused, for example, by unstable contacts, 50 random sequences of the magnetic field were applied, and a reference measurement at 0 T was also recorded between each magnetic field. The carrier concentration was deduced by $n = I_{14}B/\pi q r V_H$ where $B$ is the magnetic field perpendicularly aligned to the long axis of the NW, $q$ is the elementary electron charge and $V_H$ is the measured Hall voltage with respect to ground.

*Si NW-based IR photoconductor characterization*: The room temperature current-voltage (I-V) characteristics were measured using a Keithley 4200 semiconductor device analyzer with attoampere resolution connected to a probe station with a Faraday cage. A 1.55-µm laser was used as the excitation source in conjunction with a 50× objective lens rendering a 500-µm spot focused on top of the EBL-contacted NWs.

**Acknowledgements**
Support by the Ion Beam Center at Helmholtz-Zentrum Dresden-Rossendorf is gratefully acknowledged. Y.B. would like to thank the Alexander-von-Humboldt foundation for providing a postdoctoral fellowship. The authors also thank Lothar Bischoff and Annette Kunz for the TEM lamella preparation of individual nanowires via focused ion beam. The funding of TEM Talos by the German Federal Ministry of Education of Research (BMBF), Grant No. 03SF0451 in the framework of HEMCP is gratefully acknowledged.

**Supporting Information**

**Dynamic Monte Carlo simulations**

The development of the shape and the local composition of the nanowires during high-fluence implantation doping was modeled using the three-dimensional dynamic computer simulation TRI3DYN.[1] TRI3DYN is a 3D extension of the widely used 1D TRIDYN code.[2,3] Arbitrary 3D bodies can be arranged within a fixed cuboidal computational volume spanned by the Cartesian $x, y$ and $z$, coordinates, which are subdivided into fixed cuboidal voxels. For the current problem, an infinitely long wire was set up on a pedestal, representing the substrate with the wire axis along the $y$ direction. Periodic boundary conditions in $y$ were applied. The $x$ axis points from the top of the wire into the substrate. Laterally in $z$ direction, open boundaries were applied to simulate the irradiation of a single isolated nanowire. The total extension of the computational volume was $x_{max} \times y_{max} \times z_{max} = 260\ nm\ \times 100\ nm \times 320\ nm$ such that the spatial extension of the collision cascade fits well into the system in the $y$ direction. The voxel spacing was $2\ nm \times 2\ nm \times 2\ nm$.

Laterally uniform irradiation was applied with the ions starting at randomly selected positions on the upper $y - z\ (x = 0)$ plane at an incident energy and angle (with respect to the $x$ direction) of 60 keV and 7°, respectively. For the entry of ions at a glancing incidence (in particular at the sides of the nanowire), the voxel structure may result in an artificially enhanced entry and trapping, whereas, in reality, such ions would be preferentially reflected from the wall.[1] Consequently, an algorithm for the 3D surface planarization was developed and replaced the cuboidal surface of each surface voxel by a locally planar surface with an inclination accounting for the neighboring surface voxels. The slowing down of the ions and the development of the associated collision cascades was modeled in the binary collision approximation (BCA) using algorithms based on the stopping and range of ions in matter (SRIM).[4,5] The collisional processes altered the local composition in a number of voxels due to ion implantation, atomic relocation and surface sputtering.

In the simulation, each moving atom ("pseudoatom") represents a certain number of real atoms, which may be fractional (8.65 in the present case) and is automatically selected to minimize the computation time while retaining sufficient statistical quality of the dynamic development of the system. After a certain number of incident pseudoprojectiles (10 here), the dynamic relaxation of the system was activated. In each modified bulk voxel, the total atomic density was re-established according to the new local composition assuming fixed atomic volumes of the atomic constituents. This was accomplished by material exchange between neighboring voxels and transport from/to surface voxels (for details, see Ref. 1)

The dynamic development of the system at large irradiation fluence was significantly influenced by the surface sputtering. The choice of the corresponding parameters of the surface binding energy was non-trivial, especially in compound systems. In a default recipe, the surface binding energies of the constituents can be calculated from the actual local composition using a simple thermochemical model[6] that satisfies the balance of the heats of sublimation, the enthalpies of compound formation and the molecular binding energies of gaseous constituents. However, there is a general tendency, even for monoatomic materials, that sputtering yields are often significantly underestimated (by a factor of up to approximately 2) in BCA computer simulation.[7] Therefore, a potential adjustment of the simulation parameters based on a critical comparison of the experimental results may be considered, which, however, is often hampered by the poor availability and frequent discrepancy of experimental sputtering data. For the present system, the following procedure was adopted. Experimental sputtering data were available for Kr ions (being of similar mass as Se) incident on $SiO_2$, at, however, a significantly lower ion energy ($\leq 1\ keV$). These data can be well reproduced using TRIDYN simulation for a flat surface when reducing the surface binding energies obtained from the previous default recipe by a factor of 0.45. This choice was transferred to the present simulations apparently still leaving some uncertainty, which is difficult to quantify.